# Title: Mechanical properties of drying plant roots: Evolution of the longitudinal Young's modulus of chick-pea roots with desiccation

Pascal Kurowski[1], Christopher Vautrin[1], Patricia Genet[2,3], Lamine Hattali[4], Ramon Peralta y Fabi[5] and Evelyne Kolb[1,*]

[1]Laboratoire de Physique et Mécanique des Milieux Hétérogènes (PMMH), UMR CNRS 7636, ESPCI Paris, PSL Research University, Sorbonne Université - UPMC, Univ. Paris 06, Sorbonne Paris Cité - UDD, Univ. Paris 07, 10 rue Vauquelin, 75005 Paris, France

[2]Institut d'écologie et des sciences de l'environnement de Paris, UPMC UMR 113 – CNRS UMR 7618, 7 quai Saint-Bernard, bât A, 7[ème] étage, CC 237, 75005 Paris, France

[3]Université Paris Diderot – Paris 07, 75205 Paris cedex 13, France

[4]Laboratoire FAST, Université Paris-Sud, Université Paris-Saclay, 91405 Orsay Cedex, France

[5]Departamento de Física, Facultad de Ciencias, UNAM, CdeMx, Mexico

*Author for correspondence

Evelyne Kolb

Phone: +33 1 40 79 58 04

Email: evelyne.kolb@upmc.fr

Author e-mails:

Patricia Genet: patricia.genet@univ-paris-diderot.fr

Lamine Hattali: lamine.hattali@u-psud.fr

Pascal Kurowski: pascal.kurowski@espci.fr

Ramon Peralta y Fabi: fabi@ciencias.unam.mx

Christopher Vautrin: vautrin-christopher@hotmail.fr

**Abstract**

Mechanical characterizations of plant roots are of primary importance in geophysics and engineering science for implementing mechanical models for the stability of root reinforced-soils, as well as in agronomy and soil science for understanding the penetration of roots in soils and optimizing crop. Yet the mechanical properties of plant roots depend on their water content, which can drastically evolve with drying or flooding of the external soil. The present work deals with the determination of the longitudinal Young's modulus of single non-lignified plant roots, chick-peas (*Cicer arietinum* L.), tested in compression along their root axis for different external environments: in controlled conditions of natural drying in air or in a osmotic solution of mannitol at the isotonic concentration where no water exchange occurs between the root and the external solution.

We submitted the chick-pea radicles to successive mechanical compression cycles separated by rest periods to follow the time evolution of the root mechanical properties in drying and non-drying environments. Control experiments on non-drying roots placed in isotonic osmotic solutions showed no evolution of the root's Young's modulus whose value was around 2 MPa. On the contrary, the experiments performed in air exhibited a dramatic increase of the root's Young's modulus with the drying time, sometimes by a factor of 35. Moreover, the Young's modulus in these cases was observed to scale as a decaying power-law with the root's cross-section measured at different times of drying. We interpreted our results in the framework of the mechanics of cellular foams.

1. **Introduction**

The mechanical behavior of plant tissues, and in particular of plant roots, is an important and complex issue that raises fundamental questions as well as applied problematics in fields ranging from soil science [1], biomechanics [2] and biophysics [3,4] to food processing [5] or bio-composite engineering [6]. Yet, the mechanical approaches and knowledges in plant science are largely scattered amongst the different communities, as the mechanical forces on plant systems act on all length scales, from molecular and cellular structures up to tissues, organs, and the whole plant itself, and even up to entire communities [7].

First, mechanical characterizations of plant roots are of primary importance for agro-forestry as well as for soil science. Plant roots reinforce soils by acting mechanically as a consolidating network of bundles or fibers and by providing cohesion to the soil, therefore protecting artificial and natural slopes against shallow landslides [8] or earthquakes [9]. Over time, roots can also be responsible for considerable damages to infrastructures by developing very huge forces: Roots can crack foundations [10, 11], snap water lines, or even lift sidewalks [12]. On the other hand, the depth, architecture and morphology of the root system are known to adapt, depending on abiotic parameters like soil mechanical resistance which increases with compaction or soil drying [13]. The way a root will penetrate inside a soil, whether it will push soil grains or move them apart, be stuck or get around obstacles and follow the tortuous network of pores and fractures is not known *a priori* and raises new fields of investigation [14]. One of the first steps towards a better understanding of the root penetration mechanisms and a possible modelling of root-soil interaction is based on the knowledge of individual root mechanical properties. Therefore the present work is devoted to the determination of the rigidity of a simple root, whose water status evolves due to a drying process. In natural environment, this can often occur due to an unbalance of water potentials between soil and roots.

One of the main fields where mechanical measurements have been performed is for determining the anchorage strength of roots in a soil [15]. The root strength is a key parameter for estimating the plant resistance to overturning or lodging and it needs to be accurately characterized for obvious safety and economic reasons. The root anchorage has to counterbalance the heavy external mechanical constraints that act on the above-ground part, like wind blowing against the tree crown or through the crop canopy [16] or water flowing around riparian plants [17]. Most experimental approaches for determining the uprooting resistance of plants [18] are based on *in situ* stem pulling or bending tests but the obtained force-displacement response curves are a complex interplay between root mechanical, morphological (root diameters and their distribution) and architectural properties (root tortuosity, root branching points…) [19] as well as soil cohesion and root-soil matrix friction. Some modellings [20] have been proposed but they need to incorporate mechanical parameters like Young's modulus, bending stiffnesses and strengths of roots to predict valuable data for plant uprooting resistance, especially for tree stability under wind [21]. Yet these mechanical parameters are quite often not available for roots due to the intrinsic difficulty of measuring properties of underground materials.

Indeed, numerous studies on the mechanical properties of plant tissues have focused on stem wood. The elastic moduli along the three principal axes of wood with respect to grain direction and growth rings (longitudinal, radial and tangential) are well characterized and tabulated for various species including softwoods and hardwoods under various physiological conditions [22]. It is well established that these mechanical properties are affected by changes in moisture content below the fiber saturation point and that the stiffness and strength of wood increase (to a limit) as the wood moisture content decreases [23]. Moreover some phenomenological and world-wide correlations have been established between the mechanical parameters (like the moduli of elasticity, the modulus of rupture and the maximum compressive or shearing strengths) and the wood density for a given wood moisture content, showing the importance of controlling humidity when performing mechanical tests on plants. Other works have observed a decaying power-law relationship between tensile strength and root diameters for wood: roots with smaller diameters are observed to be stiffer and stronger [24]. According to [25], these correlations might be due to a change of cellulose composition with diameters, since smaller root diameters are often related to a larger cellulose content and therefore a higher mechanical resistance. But it is difficult to decouple what comes from a change of the biophysical properties of the roots (decrease of cellulose or increase of lignin, cell wall chemistry…) due for example to the root age [26], from a systematic mechanical trend due to a dimensional effect.

On the other hand, in the case of non-woody plants, there are relatively little works dealing with the mechanical properties of individual roots [27]. Most of the mechanical and rheological characterizations have been developed in the context of food engineering. In this field, improved knowledge is needed to maintain product quality and tissue firmness during harvesting, handling, transport and storage of processed foods like bread or fruits and vegetables. The long-term storage of tubers like potatoes [28, 29, 30] and the drying of vegetables and fruits like apples [31] or pumpkins [32] have been characterized in food technology by classical mechanical tests like compression and tensile cycles or oscillatory shear tests, at different ages or with different osmotic adjustments to control the moisture content of the plant tissue. Indeed, it is known in this community that changes in water content that lead to differences in turgor pressure at the cell scale, have a tremendous effect on macroscopic mechanical properties like elastic modulus, fracture toughness and yield strength. Some modellings have been proposed that take into account the hierarchical cellular structure of plant to explain the macroscopic mechanical tissue behavior from the microscopic scale with pressurized cells [33, 34].

This overview shows the wide scope of the mechanics of plant tissues but the information is sometimes patchy because distributed over many communities. Testing procedures for determining Young's modulus and tensile strength of roots are not always specified and parameters like strain amplitude, strain rates, root geometries and water status are not systematically given, although they could have dramatic effect on the mechanical data.

In the present work, we deal with the mechanical characterization of single non-lignified roots in controlled conditions of natural drying in air. In a first part, we will describe the model plant we used and the protocol of loading. Then we will present the evolution of the mechanical properties of the root when drying in air. We will compare them with roots immersed in mannitol solutions at the isotonic concentration. Finally we will interpret our data in the context of the cellular foams, which is a new way for rationalizing the evolving mechanical properties on single drying roots.

## 2. Material and methods

### 2.1. Plant material

As in our previous works on penetration of roots inside gaps [35], we chose Chick-pea (*Cicer arietinum* L.) as a model dicotyledonous plant because it forms a straight tap-root with a relatively large diameter of the order of one millimetre. Seeds of the variety *TWIST* from EspaceAgro were rehydrated for 2 to 4 days for germination until the radicle reached a length of around 2 cm. During this phase of germination, seeds were placed vertically in wet cotton in order to obtain straight radicles (**Fig. 1a)**.

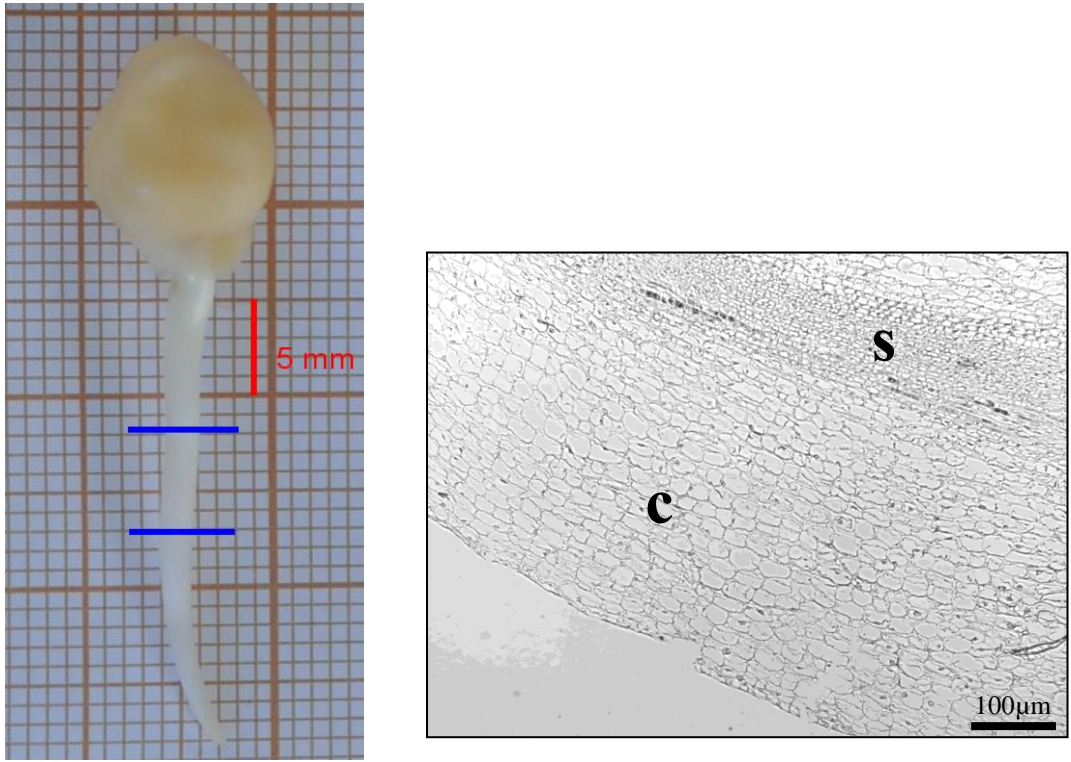

**Fig. 1**: *Chick-pea root. (a) Photo of the chick-pea seed and radicle just before sampling a fragment for mechanical characterization. The fragment is cut between the two horizontal (blue) lines. The (red) scale bar is 5 mm. (b) Microscopic observations (objective x40) of a longitudinal root section of a chick-pea radicle. The scale bar is 100 μm. The central part is the stele (noted **S**), while the outer part is the cortex (noted **C**). Only one half of the root is seen.*

A small piece of length around 5 mm (see the root fragment between the two horizontal (blue) lines of **Fig. 1a**) was cut in the central part of the root outside the tapered part of the elongation zone at a distance of around 1 cm from the root tip. That means that the fragments we investigated are located in the so-called mature (or differentiated) zone of the root, outside the elongation zone where axial growth occurs. The root is formed by a central part, the stele, surrounded by the cortex (**Fig. 1b)**. The stele represents around 1/10 of the root's diameter, while the cortex is constituted of an assembly of cell files aligned along the longitudinal axis of the root. From this microscopic section, the typical number of cells in the cortex along the diameter is around 50 (25 on each side of the stele) with a typical cell length of 30 microns along the root axis and a cell width of 15 microns along the root's diameter.

## 2.2. Compression testing

The mechanical tests were performed on the fragments of roots, placed either in ambient environment with air relative humidity ranging from 28% to 44% or in a physiological non-drying solution (mannitol bath adjusted to the osmotic status of the root at the isotonic concentration). The mechanical compression was controlled by use of an electromechanical universal testing machine (*INSTRON* model 2519-105) driven by a computer with the software Bluehill. The force was measured by use of a 10 N load cell with a precision of $2\,10^{-4}$ N. Special built-in cylindrical plates in duralumin were manufactured for compression experiments. The bottom plate was screwed inside the basis of the *INSTRON* machine, while the top plate was screwed directly below the force sensor connected to the movable part of the *INSTRON* machine (**Fig. 2**).

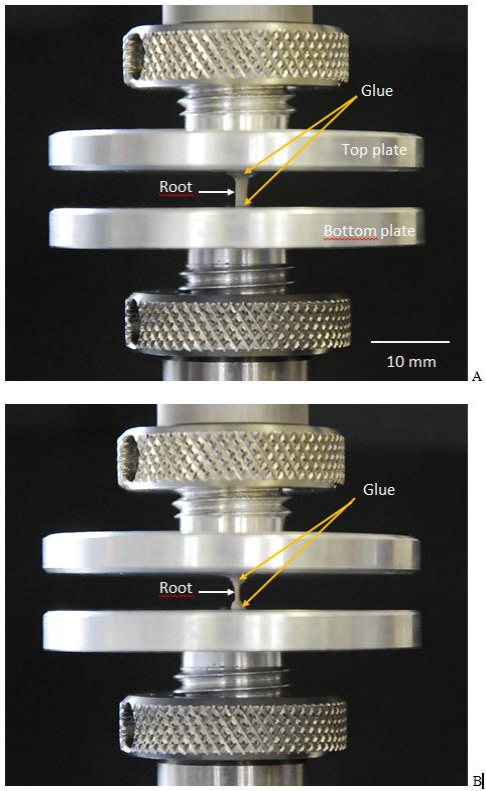

**Fig. 2**: *Front pictures of a root of initial vertical length $L_0$ = 4.7 mm placed in the INSTRON machine for compression tests. By imposing a maximum absolute strain of $|\varepsilon_{Max}|$= 1%, we thus impose the root shortening, ie. the distance of compression, to be:*

$|\Delta L_{Max}| = |\varepsilon_{Max}| \cdot L_0 = 47\ \mu m$. *As the strain rate for this experiment is $\frac{d\varepsilon}{dt}$ = 5% min$^{-1}$, the loading and unloading times are $t_L = t_U$ = 12 s.*

*The root segment is placed vertically between two parallel cylindrical plates. The bottom plate is screwed directly on the fixed part of the INSTRON machine, while the top plate is screwed to the top movable part of the*

*INSTRON machine. The two extremities of the root are fixed on the cylindrical plates by use of glue (visible on the two pictures): (a) Picture taken at time t = 15 min just before the first cycle starts. The time t = 0 corresponds to the time when the root has been cut from the seed. (b) Picture taken at time t = 153.6 min, just before the beginning of the 10<sup>th</sup> loading-unloading cycle. A significant decrease of the transverse section due to drying over time can be observed. The shoulders at the extremities of the root are the menisci formed by glue. The mean temperature during the experiment is $\theta = 25.5 \pm 1.5\,°C$, while the mean relative humidity is RH = $(39.0 \pm 2.5)\%$.*

At time *t = 0*, the small piece of length around 5 mm was cut in the mature zone of the root. The root being placed vertically, the lower extremity of the root piece was fixed on the bottom plate by means of a fast-drying cyano-acrylate glue (*LOCITE*, super glue-3 liquide, Henkel). Then the top plate was approached until it contacted the upper part of the root and a tight connection was also established by using the same glue as before. Note that cyano-acrylate glues are used in other plant biomechanical characterizations [17]. They avoid slippery boundary conditions and root radial squashing and damaging at the attachment points of the tested plant segment.

The whole experimental preparation requires duration of 13 to 22 minutes, to cut and install the root fragment in the setup and let the glue dry. After this duration, the extremities of the root were firmly attached to the plates (see below). Then the compression tests could begin. The vertical length of the tested piece of root is noted *L₀*.

Protocol of loading and unloading

The protocol consists in cycles of compression loading and unloading, each cycle being followed by waiting periods of duration $t_W$ before the next cycle occurs (see the chronogram **Fig. 3**). For each cycle we imposed a constant maximal compression strain of $|\varepsilon_{Max}| = 1\%$. We chose this small value for limiting damage and possible plastification of the root tissue. As the vertical length of the root fragment might slightly differ from one experiment to the other, we measure precisely the initial length *L₀* from image analysis of a picture taken just before the experiment. Thus we impose the compression displacement $\Delta L_{Max}$, or equivalently the root shortening to be $|\Delta L_{Max}| = |\varepsilon_{Max}| \cdot L_0$ for each cycle.

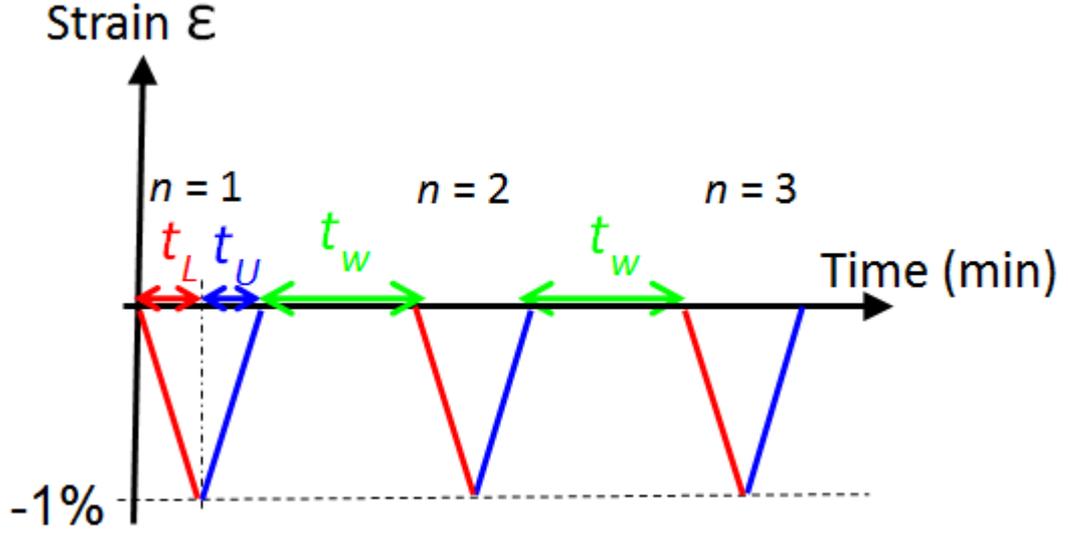

**Fig. 3**: *Schematic of the cycles of loading (loading time $t_L = \frac{\varepsilon}{\frac{d\varepsilon}{dt}}$) and unloading (unloading time $t_U = t_L$) separated by a waiting time usually set to $t_W = 15$ min. For example, for an experiment with a loading rate of $\frac{d\varepsilon}{dt} = 1\%$ min$^{-1}$, the loading or unloading times are $t_L = t_U = 1$ min.*

The loading is performed at a constant loading rate $\frac{d\varepsilon}{dt}$. The unloading part is performed just after the loading part at the same rate $\frac{d\varepsilon}{dt}$. These loading rates are ranging between 1% min$^{-1}$ and 20% min$^{-1}$ to test the possibility of viscous effects. Ten successive cycles separated by waiting periods of typically $t_W = 15$ min are performed on the same root. The protocol we used for the *INSTRON* machine is retro-controlled in displacement (for imposing a given strain) and not in force. Thus after each loading-unloading cycle, the upper part of the *INSTRON* machine comes back to its initial position and the distance between the upper and bottom plates is therefore kept constant and equals to $L_0$, the initial root fragment length.

We have verified in complementary tests that the glue has no effect on the root's mechanical measurements. For this purpose, we have tested a polymeric cylinder (rubber) of diameter 1.88 mm close to the root's diameter and of rigidity apparently similar to the ones of fresh-cut roots. First we used grips to hold the extremities of a fragment of the polymeric cylinder of length 95.38 mm in the *INSTRON* machine and determined its Young modulus in traction ($E = 13.09 \pm 0.05$ MPa) for the lowest strain rate we used at 1% min$^{-1}$. Second we used a fragment of length 5 mm (as for the root fragment) of the same polymeric material and fixed its extremities with glue with the same preparation protocol and gluing time as for roots. Then we proceeded to 12 cycling experiments at a strain rate of 1% min$^{-1}$ and a maximal compressive strain of 1%, with a waiting time of 2 minutes between each cycle of compression loading and unloading. The Young modulus obtained in compression with the glued extremities for the first cycle of compression was the same as the one in traction without any glue. Moreover the values for the following cycles were the same as the first one within an uncertainty of 2.6% over the time period of 48 minutes corresponding to the full mechanical test of 12 successive cycles. These tests on a non-biological and non-evolving material showed that the glue was dry when the mechanical tests began and did not evolve with the following cycles. The glue was sufficiently rigid for not modifying the results on the Young modulus of the fragment of polymer. We therefore rely on this protocol for measuring the Young modulus of chick-pea roots.

Measurements of Relative Humidity and Temperature

During the mechanical measurements in air, the initially fresh-cut root placed in the *INSTRON* testing machine was submitted to drying due to the surrounding ambient atmosphere and the root tissue is in an out-of-equilibrium thermo-dynamical state. This drying modifies the moisture content of the tested root and results in the shrinkage of the corresponding initially cylindrical fragment (see the differences between **Fig. 2a** and **Fig. 2b**). Consequently we continuously monitored the temperature and the humidity of the atmosphere surrounding the root during the cycling tests. These data were recorded by use of a humidity/temperature logger (*EBRO-EBI20 TH*) and stored automatically every minute during almost 3 hours starting from the time of cutting (*t = 0)* till the end of the mechanical tests.

Measurements of root diameters and cross-section with time

The drying effect was mainly visible through the change in the diameter of the root with time in the central part of the root. In order to determine the true stress $\sigma$ from the force $F$ measured with the *INSTRON*, precise measurements of cross-sections area $S(t)$ of root perpendicular to the axis of compression were performed as a function of time *t*, then:

$$\sigma(t) = \frac{F}{S(t)} \qquad \text{Eq. (1)}$$

Thus we placed a CCD camera (*NIKON* D40X with an image size of 1936 × 1296 pixels$^2$) with a 60 mm macro-lens in front of the *INSTRON* machine to capture front pictures of the root over time *t*. An image of the root in the *INSTRON* machine was taken just before the compression experiment started at time $t = t_1$. Then the next images were obtained automatically by use of the software Camera Control Pro 2 (from *NIKON*) and by imposing a regular time delay between acquisition. This period was based on the period imposed to the cycling experiments. From these images, we extracted the ones corresponding to the times $t_n$ just before the $n^{th}$ cycle began.

*ImageJ* software was then used to determine with an accuracy of 0.025 mm the diameter of the root in the middle part of its vertical length. This diameter will be called $d_F(t_n)$ (the index *F* being for Front diameter). For the last experiments, another CCD camera (*NIKON* D300S with an image size of 2144 × 1424 pixels$^2$) was placed on the lateral side of the *INSTRON* testing machine in order to capture the root profile in the vertical plane at 90 degrees from the preceding visualisation plane. A second diameter could be evaluated at the mid-length of the root and will be called $d_T(t)$.

From the estimation of these diameters over time, we compute the cross section area $S(t_n)$ at mid-length. This cross-section is rather constant over the whole length of the root except near the solidified glue menisci. For the first experiments where no transverse diameters were measured, we used the following formula assuming the root kept its cylindrical cross-section:

$$S(t_n) = \frac{\pi \, [d_F(t_n)]^2}{4} \qquad \text{Eq. (2)a}$$

For the last experiments, we used the values of the two perpendicular diameters for computing the root elliptical cross-section $S(t_n)$:

$$S(t_n) = \frac{\pi}{4} d_F(t_n) d_T(t_n) \qquad \text{Eq. (2)b}$$

These measurements of the cross-sections $S(t_n)$ at mid-lengths before each cycle *n* are necessary for determining the true stress $\sigma(t_n)$ according to Eq. (1).

Analysis of the mechanical data

The raw data from *INSTRON* provide the force $F(t)$ as a function of time $t$ and displacement $\Delta L$ of the upper movable part of the testing device, where the top extremity of the root has been attached. The data corresponding to each cycle *n* of loading-unloading are extracted (as the curves we will see later in **Fig. 7**). The usual way of determining the Young's modulus $E$ consists in extracting the initial slope $b$ in the linear part of the $F$-$\Delta L$ curve during the loading phase (equation (3)), such that:

$$\sigma = \frac{F}{S} = b\frac{\Delta L}{S} = \frac{bL_0}{S}\varepsilon \qquad \text{Eq. (3)}$$

In this formula, the strain is defined as the nominal strain $\varepsilon = \frac{\Delta L}{L_0}$, the length of the root at any time always being at least equal to 99% of its initial size $L_0$ ($|\varepsilon_{Max}| = 1\%$).

In classical tests, the Young's modulus in compression $E$ is easily determined from the slope $b$ of the loading part by identification to $\sigma = E\varepsilon$, thus leading to $E = b\frac{L_0}{S}$. As there is an evolution with the number of cycles *n* or times $t_n$, we note $b(t_n)$ the initial loading slope of the $F$ - $\Delta L$ curve for the $n^{th}$ cycle and $S(t_n)$ the root section measured just before the $n^{th}$ cycle begins; we thus obtain:

$$E(t = t_n) = b(t_n)\frac{L_0}{S(t_n)} \qquad \text{Eq. (4)}$$

However in the case of drying roots, the determination of the Young's modulus $E$ is not straightforward because we have to take into account the additional residual stress induced by root desiccation. This effect will be explicitly addressed in the *Results* part.

*2.4. Experiments in mannitol solutions*

As mentioned in the introduction, experiments conducted on plant tissues emphasise the importance of the moisture content in their mechanical behaviour. Therefore, to compare with root drying in air, we also performed experiments with roots immersed in mannitol solutions at the isotonic concentration. By definition, this is the concentration where there is no water exchange between the protoplasm of root cells and the external mannitol bath.

*2.4.1. Identification of the isotonic concentration*

First we determined the isotonic concentration of mannitol for the chick-pea root. This was done by cutting segments of chick-pea roots of 2 cm long and immersing the segments in mannitol solutions prepared in MES buffer (2-(*N*-morpholino)ethanesulfonic acid) with $pH = 5.8$ at different concentrations ranging from 0.1 M to 1 M. Then a second batch was prepared in a refined range from 0.02 M to 0.2 M. The variations of the diameters as well as the masses of the different segments were recorded as a function of time of immersion. We identified the concentration where the variations in diameters or masses were the minimum ones compared with the initial state. This gives an estimation of the isotonic concentration for chick-pea root at $c_0 = 0.14 \pm 0.01$ M. We have verified with an osmometer the corresponding osmolarity of the isotonic mannitol solution. Thus, the osmotic potential of the fresh root as defined by soil scientists or plant physiologists is $\psi = -0.34$ MPa according to Van't Hoff equation [36].

*2.4.2. Mechanical characterization of roots immersed in mannitol solutions*

A home-made plexiglass rectangular box was adapted to the *INSTRON* testing machine to mechanically probe samples immersed in baths of mannitol. The solution of mannitol with a concentration close to the isotonic one was poured into the plexiglass box before the mechanical experiment begins. The cylindrical plates were initially outside the mannitol bath so that it was possible to fix the root with glue in air. Then after a drying time of the glue of 13 to 19 minutes, the root was slowly immersed in the mannitol bath and the mechanical tests could begin.

The protocol of compression for mannitol-immersed segments is the same as for the characterization of the roots exposed to atmospheric environment. Owing to the transparent and flat walls of the plexiglass box, pictures of the root were regularly taken with the same frequency as the one used for roots tested in air.

**3. Results**

### 3.1. Desiccation of the root in air

A typical *INSTRON* experiment with 10 cycles of loading-unloading and rest periods of 15 minutes between each cycle lasts around 2.5 hours. During this time the initially fresh-cut root is exposed to atmospheric conditions and is observed to shrink laterally due to drying. The mid-length diameter $d_F(t)$ of the root observed with the CCD camera placed in front of the setup as well as the transverse diameter $d_T(t)$ are observed to decrease with time $t$. As mentioned before, the reference $t = 0$ corresponds to the time when the root has been extracted from the wet cotton and cut from the seed. The first point of measurement of root diameters is obtained around 15 minutes later, as the root had to be placed and glued in the testing machine before. During the positioning of the root in the *INSTRON* machine, a liquid meniscus of glue has formed around the two root extremities (visible in **Fig. 2**). When the mechanical test begins, these menisci have solidified and anchor the root extremities. The section at root mid-length $S$ is computed from equations (2) and plotted as a function of time $t$ in **Fig. 4**. This section $S(t)$ is observed to decrease rapidly with time and eventually stabilises at a constant value $S_\infty$ depending on the drying process. The continuous lines in **Fig. 4** are exponential decaying fits with the following expression:

$$S(t) = (S_0 - S_\infty) \exp\left(-\frac{t}{\tau}\right) + S_\infty \quad \text{Eq. (5)}$$

$S_0$ is the extrapolated value of section at time $t = 0$. The time $\tau$ is the typical time constant of the drying process. It is of the order 1 hour but varies amongst experiments performed with different aspect ratios (root's diameter relative to its length) and environmental conditions (relative humidity and/or temperature).

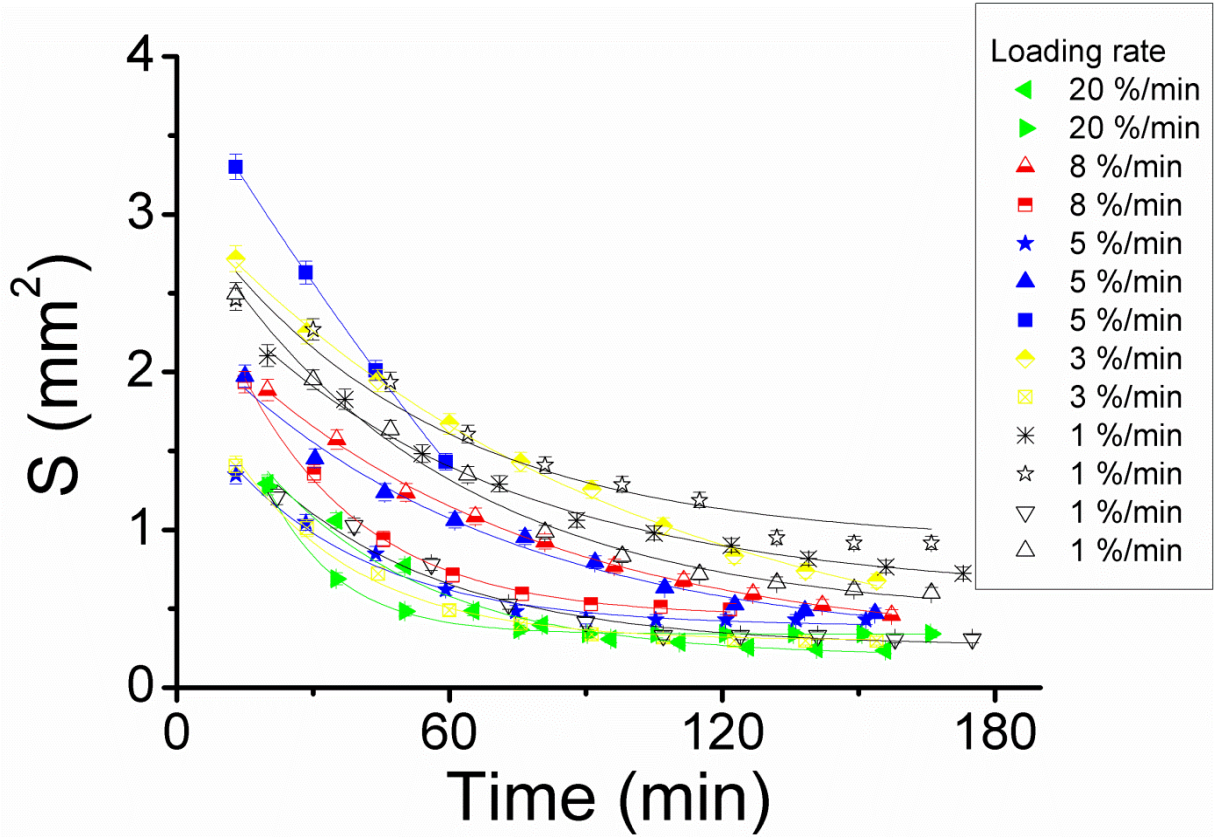

**Fig. 4**: *Evolution of the section S at mid-length of the root drying in air as a function of time t since the root has been cut from the seed and for different loading rates. The symbols are obtained from experimental measurements of the root diameters according to equations (2) for the successive compression cycles, the lines are exponential decaying fits according to equation (5). One of the root experiencing a loading rate of 5% min$^{-1}$ broke between cycles n = 4 and n = 5.*

This time $\tau$ extracted from the individual fits of $S(t)$ is plotted in **Fig. 5** as a function of the equivalent diameter $d_{ss}$, which characterizes the importance of bulk (root volume $V_{root}$) relative to the air-exposed outer surface of the root ($A_{root}$). Indeed we chose to define $d_{ss}$ as:

$$d_{ss} = 4 \frac{V_{root}}{A_{root}} \qquad \text{Eq. (6)a}$$

Note that if the root was a perfect cylinder of length *L* and diameter *d*, then $d_{ss}$ would be equivalent to the root's diameter *d*, as:

$$d_{ss} = 4 \frac{\pi d^2 L/4}{\pi d L} = d \qquad \text{Eq. (6)b}$$

**Fig. 5** shows that the timescale $\tau$ is observed to increase with $d_{ss}$, as expected from a drying process involving the surface exposed to air (where evaporation proceeds) relative to the reservoir of water in the root bulk. The slender the root is, the faster the drying process is. Moreover the symbol sizes (also represented with coloured code) in **Fig. 5** are proportional to the relative humidity *RH*: this representation shows that the lower (respectively higher) $\tau$ value corresponds to the smaller (resp. higher) *RH*, as the drying process is more efficient for lower relative humidities. But the tendency is not very clear as *RH* does not vary much between the different experiments.

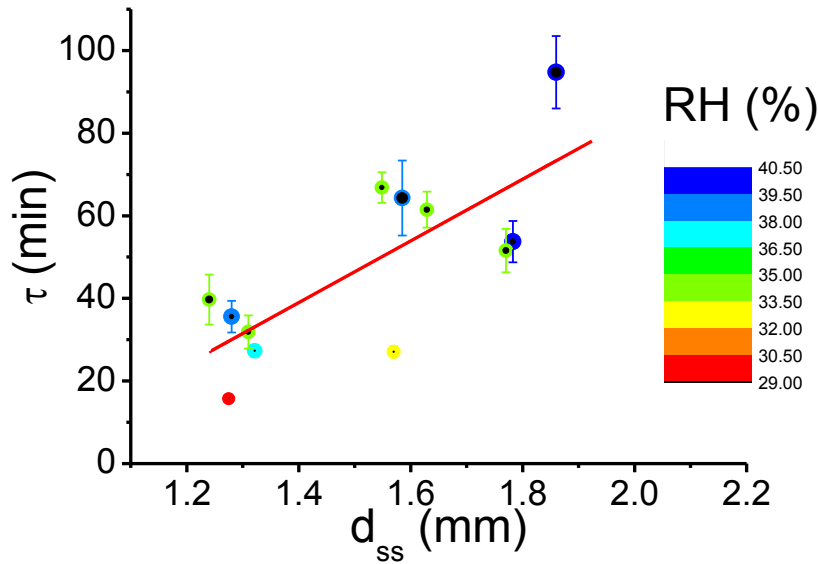

**Fig. 5**: *Drying timescale $\tau$ of a segment of root placed in the INSTRON machine, obtained from the fits of **Fig. 4** with the equation (5) as a function of the equivalent diameter $d_{SS}$ defined by equations (6). The error bars correspond to the standard errors on $\tau$ obtained from the fits. The different values of the relative humidity RH are proportional to the sizes of the circles in the figure (and are also represented by use of a color map). The straight line is a linear fit (adjusted $R^2 = 0.52$).*

The measurements of $d(t)$ or $S(t)$ are used further for computing the true stress experienced by the root during the cycling processes.

### 3.2. Force-displacement curves and Young's modulus of root drying in air

<u>Evolution of force with time</u>

From the *INSTRON* testing machine, we monitored as a function of time *t* the force *F(t)* experienced by the root as well as the displacement $\Delta L(t)$ imposed to the upper side of the root, starting from a root length $L_0$. The typical signal of $F(t^* = t - t_1)$ is plotted in **Fig. 6** over the whole duration of the experiment for the root in air. The first cycle starts at $t = t_1$ or equivalently at $t^* = 0$. The 10 compression loading-unloading cycles are clearly visible on **Fig. 6** by the huge and sharp drops at regular timescales (noted from 1 to 10). However we observe a smooth increasing signal superimposed on these 10 cycles. This increasing envelope of $F(t^*)$ corresponds to the development of a residual traction force with time. This shift is also observed in test experiments where the root is maintained at a constant length $L_0$ in the *INSTRON*, without being submitted to any loading-unloading cycles. This evolution of the force is a manifestation of the drying residual stress. It is due to the traction exerted by the drying root that would also shrink longitudinally (along the vertical axis) if it would not be maintained at a fixed length $L_0$ between each cycle.

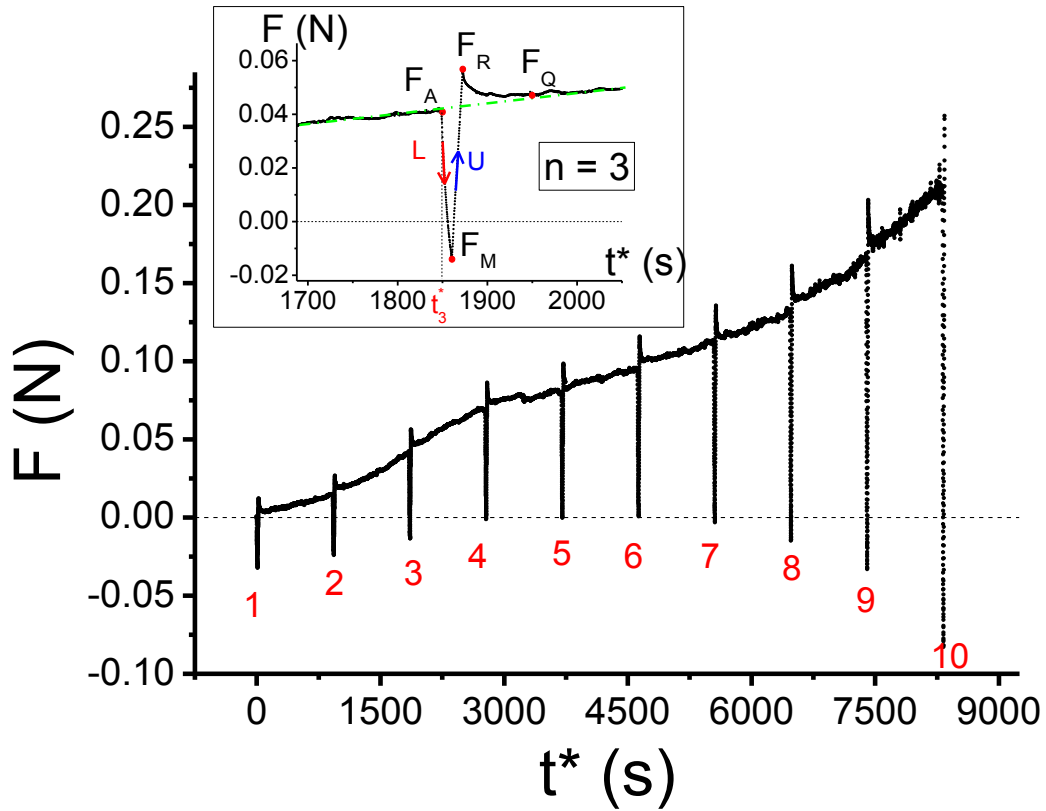

**Fig. 6**: *Force detected by the INSTRON machine as a function of time $t^* = t - t_1$, with $t_1 = 15$ min. A positive (negative) value of the force corresponds to a traction (respectively compression). The numbers below the main curve correspond to the different cycles from n = 1 to n = 10. The insert is a zoom around the cycle n = 3.*

The inset of **Fig. 6** is a magnification of the preceding curve $F(t^*)$ around the cycle $n = 3$. The loading part of the compression cycle starts at time $t_3^* = t_3 - t_1$ at a force $F_A \neq 0$ and corresponds to the steep decaying part of the curve *F(t)* (*L* (red) arrow) till the force value $F_M$. The sign of $F_M - F_A$ is negative because it is a compression. Then the unloading part (decompression) follows with an increasing and steep signal (*U* (blue) arrow) and reaches the force $F_R$ at the end of the cycle. We observe that $F_R \neq F_A$, meaning that there is a small plastic or viscous component induced by the cycling. However once the cycle is completed, there is a relaxation of the force $F_R$ till a value $F_Q$ over a timescale $\tau_{Relax}$ of the order of 1 min for this experiment. Note that this force relaxation is observed after every cycle. After this relaxation, the force value $F_Q$ almost reaches $F_A$, the value that it had before the cycle starts. The small difference between $F_Q$ and $F_A$ is due to the continuing drying process between $t = t_3$ and $t = t_3 + 2 \cdot \frac{|\varepsilon_{Max}|}{\frac{d\varepsilon}{dt}} + \tau_{Relax}$, that leads to a regular increasing force (dash-dotted (green) line in **Fig. 6 inset**) with a gentle slope. This observation is in favour of a viscous relaxation, with the typical timescale $\tau_{Relax}$ probably due to a poro-elastic process (water transport through the elastic tissues of the cell walls) [37]. Thus the difference between $F_R$ and $F_A$ might be due to a viscous effect rather than a plastic component.

Force-displacement curves

From the signal *F(t)*, we have extracted the sequence of data corresponding to each of the 10 cycles. A typical force *F* - displacement Δ*L* for the first cycle *n* = 1 is presented in **Fig. 7a**. We recall that a negative force corresponds to a compression force. During the loading phase (***L*** (red) arrow in **Fig. 7a**), the absolute value of the force increases till 32 mN while the distance between the fixed extremities of the root decreases from $L_0$ to $L_0 - |\Delta L|$. We recall that the maximum |Δ*L*| is prescribed by the measured value of the initial root length $L_0$ and by the imposed maximum strain of $|\varepsilon_{Max}|$ = 1% , meaning the imposed maximum displacement is |Δ*L*| = 1% × $L_0$ = 47 µ*m* for this particular experiment where $L_0$ = 4.7 mm. Although the maximum compressive strain is very small, the first loading curve is not linear over the full range of deformation and a linear relationship between force and displacement with a constant slope $b_L = \frac{\Delta F}{\Delta L} > 0$ (as Δ*F* < 0 and Δ*L* < 0) is only observed in the first 10 microns of compression, corresponding to an absolute strain lower than 0.2%. The unloading phase (***U*** (blue) arrow in **Fig. 7a**) is characterized by a larger linear range with a slope $b_U$ smaller than $b_L$. As already mentioned from the observations of **Fig. 6 inset** , a small hysteresis is observed between the loading and unloading curve with a positive residual force $F_R$ after unloading, when the upper part of the *INSTRON* machine comes back to its initial position $L_0$.

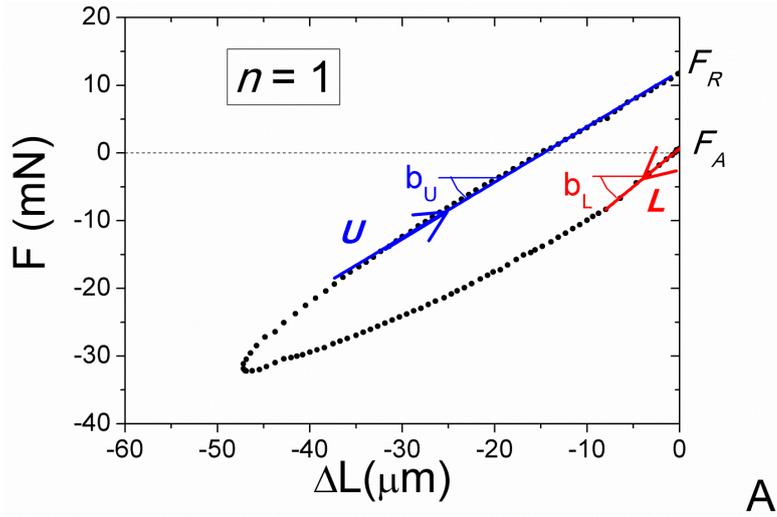

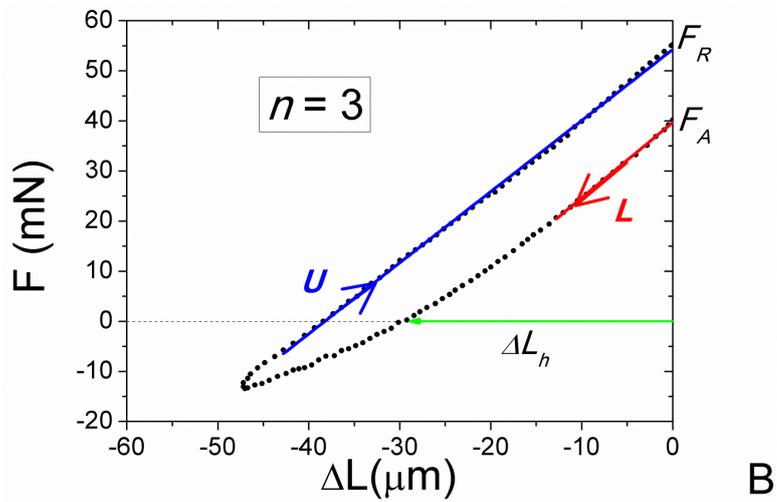

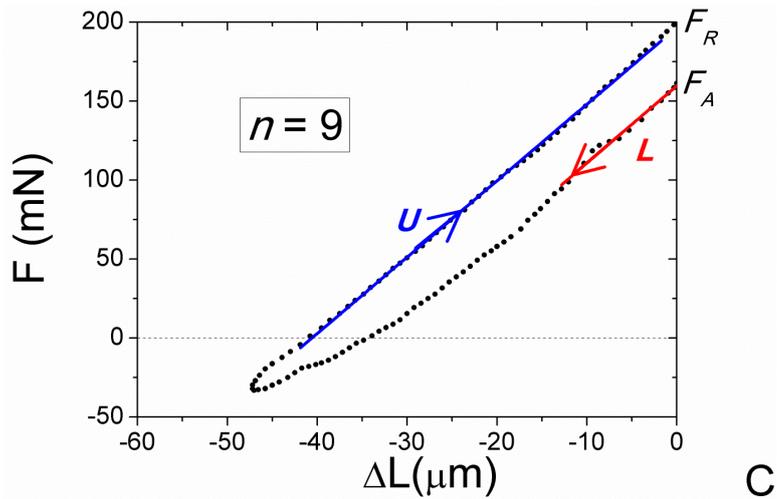

**Fig. 7**: *Compression force versus displacement for the first (n = 1) (a), third (n = 3) (b) and ninth (n = 9) (c) cycles at a loading rate of 5% min$^{-1}$ for a fragment of a four-day old radicle. The mechanical experiment has been performed with a mean temperature θ = (25.9 ± 1.4) °C and a mean relative humidity RH = (38.8 ± 1.8) %. Note the change of the vertical scale range with n.*

The following cycles are also presented in **Fig. 7b, c**. Due to the drying residual stress that increases with time of drying, the force value ($F_A$) at the starting point of the cycle is shifted towards positive force value (traction force) and increases with the number $n$ of the cycle (or possibly saturates for other experiments, depending on the root drying process). As for $n = 1$, we define the slopes $b_L(t_n)$ and $b_U(t_n)$ for the loading and unloading phases respectively of the $n^{th}$ cycle that begins at time $t = t_n$. For this experiment we observed that the relative difference between these slopes is smaller than 10 % as soon as $n > 1$. We observe the same trend for other roots.

From these slopes and from the measurements (**Fig. 4**) of the root section $S(t_n)$ that varies due to drying, it is in principle possible to determine the Young's modulus according to equation (4). But the analysis has to take into account the additional effect of drying residual stress. Namely at time $t = t_n$, just before the $n^{th}$ cycle begins, there is an additional tensional strain induced by the longitudinal root's drying: the reference length $L_{Ref}$ of the non-stressed (relaxed) root is not the initial length $L_0$ at the beginning of the $n^{th}$ cycle but it has a smaller value. An estimation of $L_{Ref}$ can be obtained by identifying the abscissa $\Delta L_h$ (see **Fig. 7b**) of the force $F$ - displacement $\Delta L$ curve where the force passes through zero:

$$L_{Ref} \approx L_0 + \Delta L_h = L_0 - |\Delta L_h| \text{ as } \Delta L_h < 0$$

And the corresponding Young's modulus (the one based on the slope of the loading part $b_L(t_n)$) follows as:

$$E_L(t = t_n) = b_L(t_n) \frac{L_{Ref}}{S(t_n)} \approx b_L(t_n) \frac{L_0 - |\Delta L_h(t_n)|}{S(t_n)}$$

Note that whenever $\Delta L_h(t_n)$ can be measured, its absolute value is at maximum equals to $|\Delta L_{Max}| = 1\% \times L_0$, which is always much smaller than $L_0$, such that $L_{Ref} \approx L_0$. Even in the cases where $\Delta L_h(t_n)$ is not measurable (the loading curve does not intersect the axis $F = 0$), an estimate of this value is given by $|\Delta L_h(t_n)| \approx \frac{F_A(t_n)}{b_L(t_n)}$. From the measurements of $F_A(t_n)$ and $b_L(t_n)$, we have verified that it is always smaller than $1.5\% \times L_0$. This means that the absolute value of the residual strain induced by the longitudinal drying is smaller than 1.5%.

Therefore the formula of equation (4) is recovered and we chose to define the longitudinal Young's modulus of the root with:

$$E(t = t_n) \approx b_L(t_n) \frac{L_0}{S(t_n)} \qquad (\text{Eq. 7})$$

In the same way we can define the Young's modulus $E_U(t = t_n)$ based on the unloading slopes $b_U(t_n)$. The two values are reported in **Fig. 8** as a function of time since the root has been cut. In the error bars, the values $E$ and $E_U$ are very similar and they are observed to increase dramatically with time. Thus, in the following we will only use the $E$ value and investigate it systematically as a function of time for all the experiments.

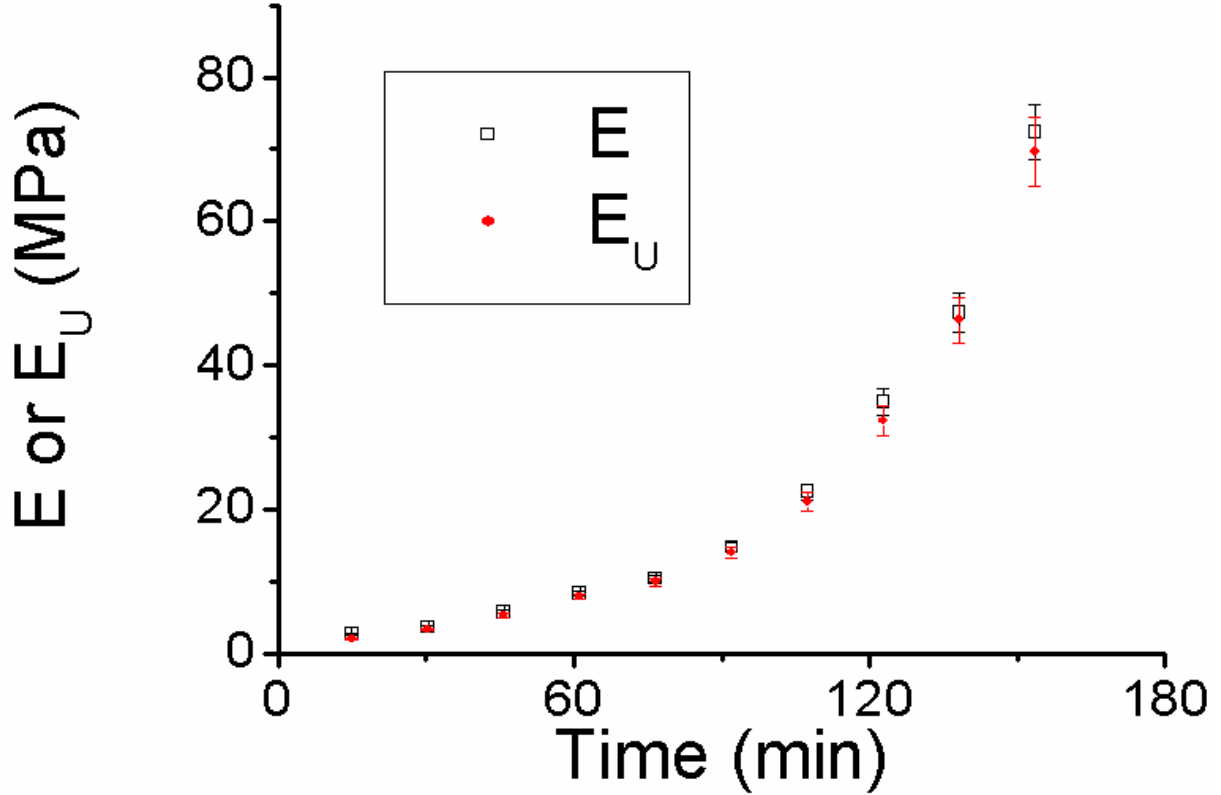

**Fig. 8**: *Two determinations of the Young's modulus as a function of time for an experiment with a loading rate of 5% min$^{-1}$. The (black) open squares: method 1 based on the initial loading slope of the force-displacement curve. (Red) full circles: method 2 corrected for the drying residual stress and based on the unloading slope of the force-displacement curve.*

Evolution of the Young's modulus with time

**Fig. 9** shows the evolution of the Young's modulus $E$ as a function of time $t$ for different roots and different strain rates. The data are largely scattered and no obvious dependency with strain rates is observed. The striking feature of **Fig. 9** is the dramatic increase of $E$ with the time of drying $t$. For example, $E$ increases from $E_1 = 8.0 \pm 0.3$ MPa at time $t_1 = 13$ min to $E_{10} = 285 \pm 20$ MPa at time $t_{10} = 154$ min for the upper curve ((yellow) cross center square in **Fig. 9**) performed at a strain rate $\frac{d\varepsilon}{dt} = 3\%$ min$^{-1}$: this gives an increase of Young's modulus of more than a factor of 35 for this first root. A similar test with the same strain rate on a second root ((yellow) full half up diamond in **Fig. 9**) leads to an increase of $E$ from $E_1 = 4.1 \pm 0.1$ MPa at time $t_1 = 13$ min to $E_{10} = 66 \pm 4$ MPa at time $t_{10} = 154$ min, corresponding to an amplification factor of 16. The first root with the larger increase of Young's modulus (factor of 35) corresponds to an initially thinner root ($d_F(t_1) = 1.490 \pm 0.025$ mm) compared with the second root ($d_F(t_1) = 1.860 \pm 0.025$ mm). Consequently and according to the results of **Fig. 5**, the first root has a smaller drying timescale: $\tau = 27.3 \pm 1.5$ min, while $\tau = 95 \pm 8$ min for the second root, meaning that the faster drying process is associated to the larger Young's modulus increase. All the curves in **Fig. 9** show an increase of $E$ with time, but some exhibit an inflection point with a saturation of $E$ at large times, while others do not. The Young's modulus for the first cycle amongst the different experiments is ranging from $E_1 = 1.4$ MPa to $E_1 = 9.2$ MPa and its mean value is 3.9±2.3 MPa (with a median value of 3.1 MPa).

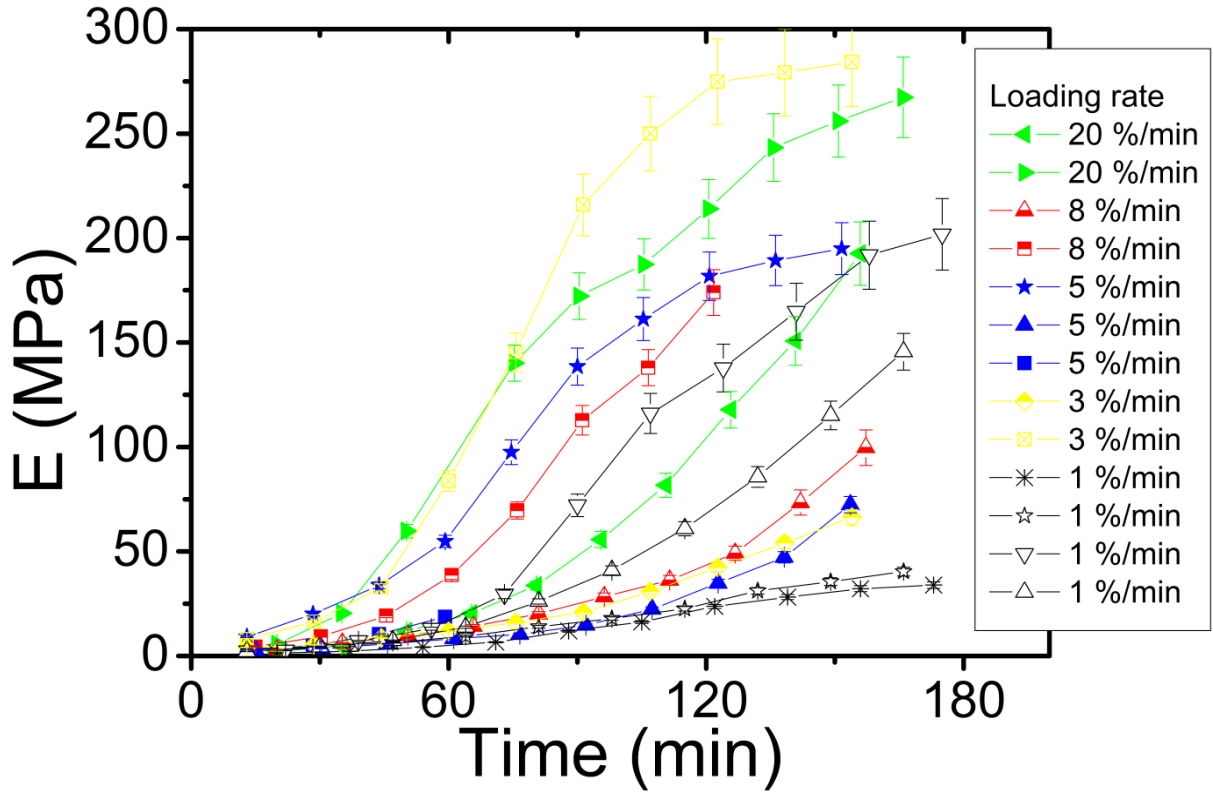

**Fig. 9**: *Evolution of the Young's modulus as a function of time for different loading rates for segments of roots submitted to loading-unloading cycles of compression in air environment. The symbols (and color code) are the same as in **Fig. 4**. One of the root experiencing a loading rate of 5% min$^{-1}$ broke between cycles n = 4 and n = 5.*

### 3.3. Experiments in mannitol solutions

The same experiments have been performed with segments of roots immersed in mannitol baths at concentrations close to the isotonic one $c_0 = 0.14 \pm 0.01\ M$. **Fig. 10** shows the evolution of the Young's modulus as a function of time since the root has been cut from the seed. Note the vertical scale in **Fig. 10**: it is 50 times smaller than in **Fig. 9** for roots drying in air over the same duration. The Young's modulus only varies by maximum a factor of 2 ((red) full circles in **Fig. 10**). Compared with roots exposed to air, it is clear that the Young's modulus of roots placed in a mannitol bath at the isotonic concentration does not almost evolve with time.

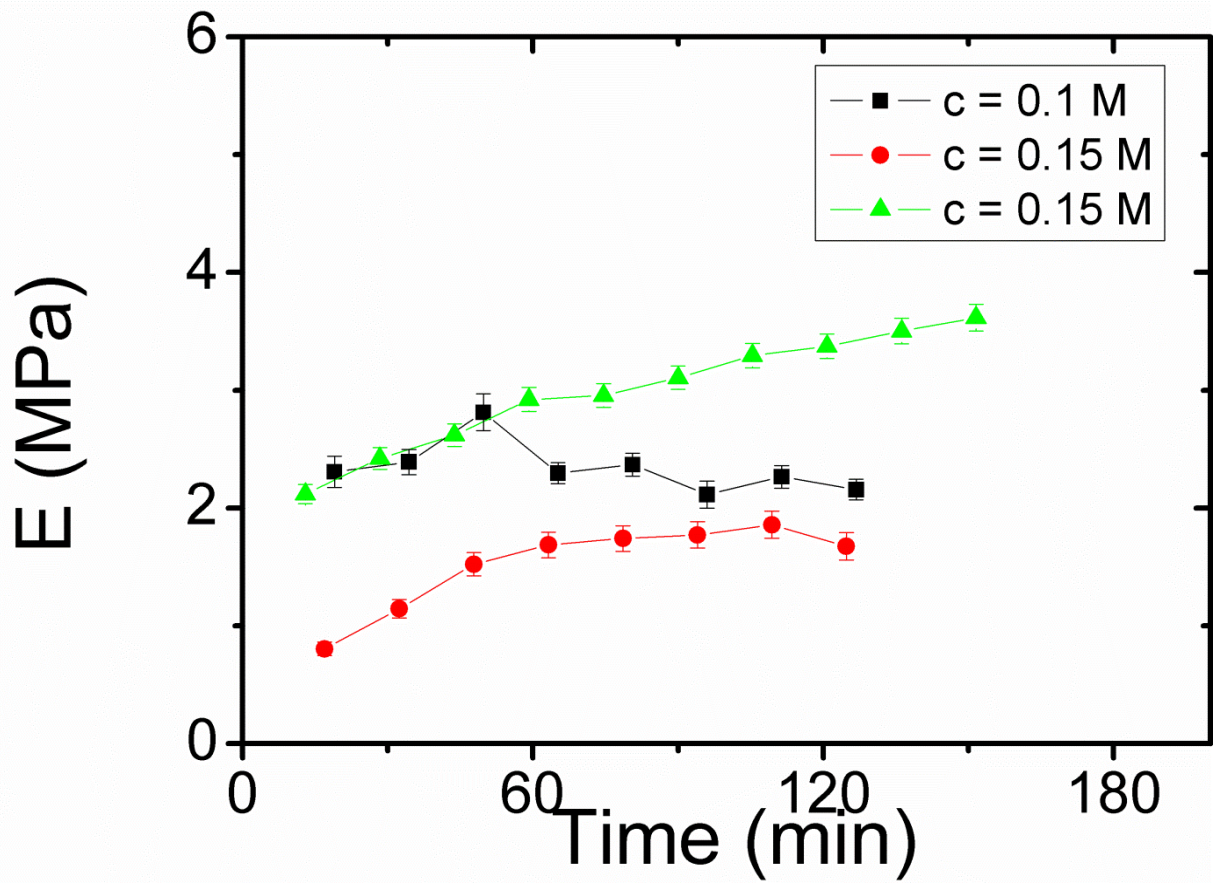

**Fig. 10**: *Evolution of the Young's modulus as a function of time for roots immersed in different concentrations of mannitol close to the isotonic one. The roots are submitted to compression cycles separated by waiting times with a protocol similar to the roots drying in air. The loading rate during the compression cycles was 5% $min^{-1}$.*

**Discussion**

**4.1. Young's modulus as a function of the root section**

We observed in **Fig. 9** the scattering and the huge increase of the Young's modulus with the time of root drying in air. No similar increase is observed for the roots experiencing the same protocol of compression cycles, but immersed in mannitol solutions at or close the isotonic concentration (**Fig. 10**). These results indicate that the drastic Young's modulus increase with time for roots placed in air is due to the drying process and is not due to a possible cumulative damage produced by the 10 successive cycles. On the other hand, it is known that tissue damage could release ethylene, a volatile plant hormone [38], which would modify the cell wall properties into the plant tissues and might result in an evolution of the root's Young modulus with time. Although we have not analysed the ethylene emission, an evolution of the mechanical properties was not observed in the root fragments immersed in mannitol even they have been excised from the tissue and submitted to 10 loading-unloading cycles. These results suggest that the role of ethylene is probably negligible in our experiments.

In food processing, the elastic modulus of a tissue is known to change with time, stress, physiological status and the dimensions of the tissue sample [28-32]. In soil science, there has been observed a dependency of the Young's modulus with the root diameters amongst different root types, ages or locations from the apex [24-26]. In particular, in numerous works, the Young's modulus is observed to exhibit a phenomenological power-law decrease with the root diameters. In our case, the root diameters and therefore the root section is varying with time for the **same** root drying in atmospheric environment. As the drying process is more or less rapid depending on the root aspect ratio ($d_{ss}$), the good control parameter is not the time but the cross-section ($S$) for a given length $L_0$. Therefore we plot $E$ as a function of the transverse section $S(t)$ at the mid-length of the root. **Fig. 11** shows $E(S)$ in log-log scale for the same experiment as in **Fig. 8**. The Young's modulus $E$ is observed to increase with decreasing $S$ as a power-law with an exponent of $\alpha = 2$. A departure from this power-law is observed for small $S$ values, when $S/S_1 \leq 0.26$, meaning for root diameters that have decreased by a factor larger than 2.

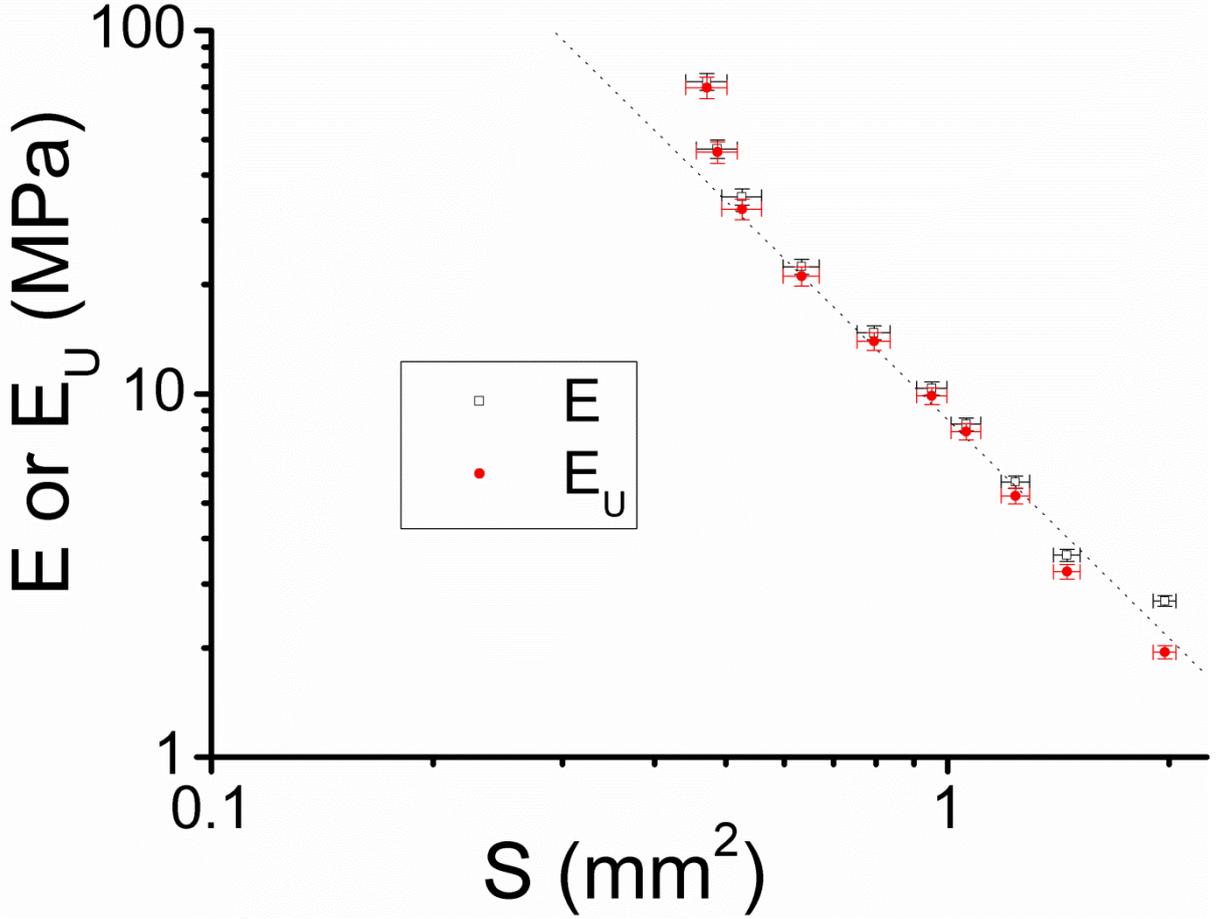

**Fig. 11:** *Two determinations of the Young's modulus as a function of the section S(t) for an experiment with a loading rate of 5% min$^{-1}$. The (black) open squares: method 1 based on the initial loading slope of the force-displacement curve. (Red) full circles: method 2 corrected for the drying residual stress and based on the unloading slope of the force-displacement curve. The dotted line is a guide for the eye corresponding to y = 8.5 x$^{-2}$, with a slope of -2.*

**Fig. 12** presents in log-log scales the rescaled data $E/E_1$ versus $S/S_1$ for all experiments, where $E_1$ and $S_1$ are respectively the Young's modulus and transverse section obtained for the first cycle of loading. By this rescaling, we obtain a better data collapse compared with **Fig. 9**, namely there exists a range where a common behaviour is observed: the ratio $E/E_1$ scales like a power-law as a function of $S/S_1$.

$$\frac{E}{E_1} \propto \left(\frac{S}{S_1}\right)^{-\alpha} \quad \text{Eq. (8)}$$

The two (blue) dashed lines correspond to a slope of -2. For comparison, the slopes of -1 and -3 have also been represented in **Fig. 12**. The exponent $\alpha$ seems to be closer to $\alpha \approx 2$ for most experiments, although the range observed for this power-law is rather limited. For larger times, i.e. for smaller $S/S_1$ values, there is again a departure from this power-law, indicating a change of behaviour.

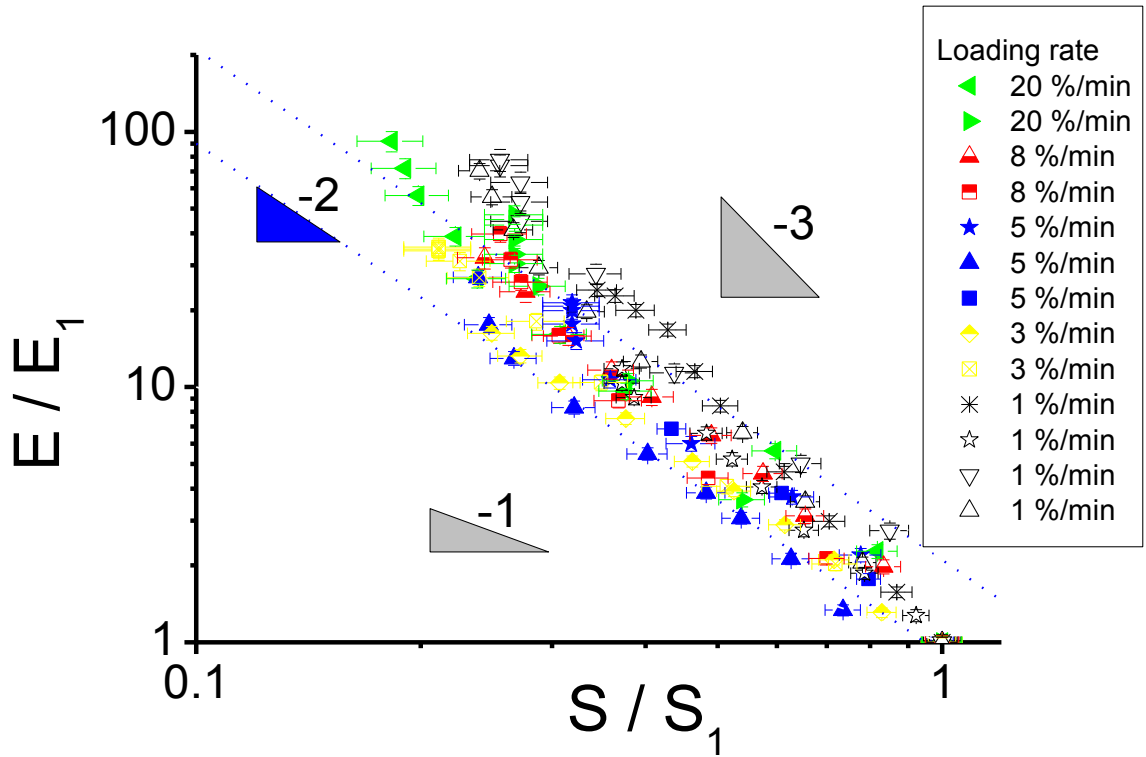

**Fig. 12**: *Rescaling of the Young's modulus E as a function of the root section S for the different experiments in log-log scales. The Young's modulus E(t), as well as the section S(t) which depend on time have been normalized by their corresponding values $E_1$ and $S_1$ at time $t = t_1$. The two dotted (blue) lines are guides for the eye and correspond to power-laws with exponent -2. For comparison, the slopes of -1 and -3 are also represented by the slopes of the gray triangles.*

### 4.2. Modelling

The longitudinal Young's modulus $E(t)$ of a root exposed to air and compressed along its axis is observed to scale with its section $S(t)$ at mid-length at different times $t$ of drying. This behaviour is reminiscent of what is observed in solid foams or engineering cellular solids [39]. We briefly recall the main results observed in literature for engineering cellular solids and mentioned how it has been applied in material science for food engineering. Then we propose a new adaptation of this model for the case of drying roots, whose mechanical properties evolve with time.

Solid foams or engineering cellular solids

The macroscopic Young's modulus ($E$) of cellular solids is related to the Young's modulus of the solid part ($E_S$) forming the cell wall material and to its relative density [40], which gives in the simplest versions the following relation:

$$\frac{E}{E_s} = C \cdot \left(\frac{\rho}{\rho_s}\right)^m = C \cdot \phi^m \qquad \text{Eq. (9)}$$

where $\rho$ is the density of the cellular solid, and $\rho_s$ is the density of the cell wall material. The relative density $\frac{\rho}{\rho_s}$ is equivalent to the volume fraction $\phi$ of the solid part.

The parameters $C$ and $m$ are constant. The value of $C$ is of the order of 1 for equi-axis cell and is related to the cell geometry. The value of $m$ depends on the type of cellular solids open or closed cell foam (see **Fig. 13a** and **Fig. 13b** for cubic cells adapted from the book of Gibson and Ashby [41] for better visualization) and the way the cellular solid is loaded [42]. For example, the cell geometry of the open-cell foam of **Fig. 13a** (solid part only at the edges of the polyhedra forming the cell) with typical cell size $a$ and cell edge thickness $b$, gives a solid fraction $\phi \approx \left(\frac{b}{a}\right)^2$ for low relative density [41]. The corresponding exponent of equation (9) will be $m = 2$, because the foam deformation at low strain is dominated by the bending of the cell edges (**Fig. 13b**), such that $\frac{E}{E_s} \approx \phi^2 \approx \left(\frac{b}{a}\right)^4$. In the case of closed cell foams (**Fig. 13c**), the exponent of equation (9) will be $m = 3$ for low strain [43]. Then the deformation is dominated by the bending of the faces (plates) instead of the bending of the edges (beams). In all these cases, a direct relationship can be established between the macroscopic mechanical properties at the level of the tissue ($E, \rho$) and the microscopic properties at the cell wall level ($E_s, \rho_s$).

Solid foams in plants

According to Gibson [34], these cellular models can be applied to describe the mechanical behaviours of plant materials like the simple tissues of parenchyma in fruits and root vegetables which have a hierarchical structure. Then the parenchyma is made of thin-walled, polyhedral cells (primary cell walls), which are densely packed and resemble an engineering closed cellular foam. Closed-cell foams refer to the geometrical structure of the plant cells with solid walls spanning the faces of the polyhedral cell (like in **Fig. 13c** for cubic cell). This is the case for example for apples or potato tubers which are nearly entirely parenchyma cells.

Indeed, experimental measurements of Young's modulus of these kinds of tissues exhibit power-law dependencies with the solid volume fraction (calculated without the protoplasm filling the cells), ie. $\frac{E}{E_s} \approx \left(\frac{\rho}{\rho_s}\right)^m \approx \phi^m$ like in equation (9) for engineering foams. Depending on the turgor pressure, different exponents *m* are observed [43], as explained later.

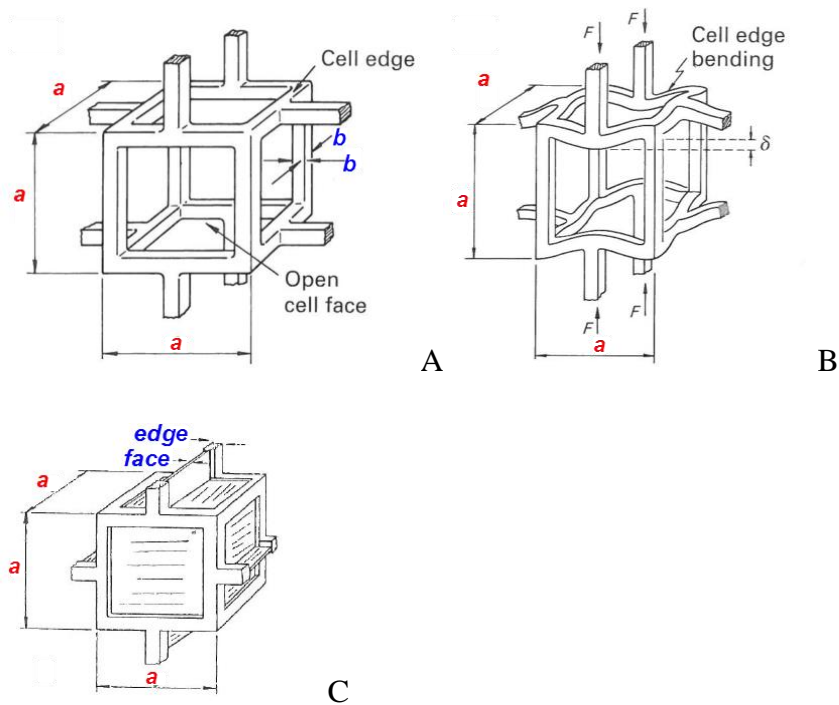

**Fig. 13**: *Adapted from the book of [39]. (a) Schematics of an open cell foam with a cubic elementary cell. The cell has a typical size a, the material is concentrated at the edges (beams) with a typical thickness b. (b) Schematics of the deformation of the cubic cell by bending of the edges, resulting from an uniaxial compression of the foam tissue along the vertical axis. (c) Schematics of a closed cell foam with material along the edges and the faces of the cubic cell.*

Root tissue interpreted as a solid foam: Adaptation of the cellular model to our case

In our experiments, the tissue under consideration is a fragment of a chick-pea radicle which clearly has a more complex, heterogeneous and anisotropic structure compared with the parenchyma of apples or potatoes. However we observed a scaling law relating the root's Young's modulus to its time decreasing section. The root structure is multicellular with around 50 cortical mature cells in the root cross-section (see **Fig. 1b**). Thus we developed approaches similar to the ones summarized in the preceding section and adapt them in a new way for mechanical properties evolving with drying time.

As the root length $L_0$ is kept constant throughout the experiment, the experimental power-law obtained in equation (8) can be reformulated as following:

$$\frac{E(t)}{E_1} \propto \left(\frac{L_0 S(t)}{L_0 S_1}\right)^{-\alpha} = \left(\frac{V(t)}{V_1}\right)^{-\alpha} \qquad \text{Eq. (10)}$$

where $V(t)$ is the total root volume at time $t$ and $V_1 = V(t_1)$ its value at time $t = t_1$.

The observed macroscopic decrease of the section $S(t)$ with time of drying is linked to the lateral shrinkage of the root and leads to a radial compaction of the root tissue, which we will consider further as a solid foam. Thus the solid fraction $\phi(t)$ of the cell walls in the tissue increases with time $t$ and according to equation (9) the Young's modulus $E(t)$ of the tissue should also increase. More precisely, we can express the ratio $\frac{E(t)}{E_1}$ between the Young's modulus of the tissue $E(t)$ at time $t$ and the one $E_1$ for time $t_1$ just before the first cycle, by introducing $E_S$, the Young's modulus of the solid part forming the cell wall material. If we assume that the solid part (total amount of cellular walls) keeps its volume ($V_S(t) = V_S(t_1) =$ Constant) during the drying process in air as well as its mechanical properties ($E_S(t) = E_S(t_1) =$ Constant), it follows that the expected ratio $\frac{E(t)}{E_1}$ from the theory of cellular foams should scale like:

$$\frac{E(t)}{E_1} = \frac{E(t)}{E_S(t)} \cdot \frac{E_S(t_1)}{E_1} \propto \left[\frac{\phi(t)}{\phi(t_1)}\right]^m = \left[\frac{V_S(t)}{V(t)} \cdot \frac{V_1}{V_S(t_1)}\right]^m = \left(\frac{V(t)}{V_1}\right)^{-m} \qquad \text{Eq. (11)}$$

Under the above mentioned hypotheses and by using the simple scaling argument $\frac{E(t)}{E_S} \propto [\phi(t)]^m$, we recover a theoretical scaling law in equation (11) similar to the experimental one of equation (10). This rationalizes the scaling law $\frac{E(t)}{E_1} \propto \left(\frac{S(t)}{S_1}\right)^{-\alpha}$ observed experimentally in **Fig. 12** except for very large times of drying in air. Furthermore the identification of the exponents in both equations (10) and (11) leads to: $m = \alpha \approx 2$, as if the drying roots in air were behaving like dry open-cell foams.

Is the "dry open-cell foam" behaviour of the drying root tissue plausible?

First we focus on the question of dry foams. We have mentioned previously that the exponent *m* of $\frac{E}{E_s} \approx \left(\frac{\rho}{\rho_s}\right)^m \approx \phi^m$ depends on the turgor pressure in the models developed by [33, 34, 43]. In this paragraph, we recall the main results obtained with these models. At normal or high turgors, the interpretation of [41] is that the plant cells are filled with liquid, the cell walls are taut and the deformation of the plant tissue is dominated by the stretching of the faces of the cell walls. This results theoretically in an exponent *m* = 1 and a rather linear stress-strain curve over the full elastic regime. These behaviours are observed experimentally for various plants like potatoes or apples at high turgor [28-31]. At low turgor pressure, the mechanisms explained by these cellular models are the following: the cell walls are no longer taut and the initial deformation of the tissue in compression is dominated by bending instead of stretching, such that the initial Young's modulus is smaller than the one for high turgor. For example, for fluid-filled closed cell foam, this results in an exponent *m* = 3 as it was observed on carrot tissues [43]. But when the uniaxial deformation increases, the filling fluid's volume is finally felt and the Young's modulus is then dominated by the cell-wall stretching and an exponent *m* = 1 is recovered. This change of exponent with increasing strain explains the non-linear form of the stress-strain curves for low turgor even in the elastic regime. The transition depends on the cell water filling and is therefore related to the turgor pressure. These effects are observed experimentally for different plant tissues, when immersed in various osmotic baths for controlling their water status [44], and thus, their degree of desiccation.

For our experiments on root tissues, the first cycle of loading begins 13 to 22 minutes after the root has been cut from the chick-pea seed. Meanwhile the root exposed to air probably partly loses its turgor and the root tissue might be considered as a dry foam [33], at least in the first part of the loading curve where the strain is small ($|\varepsilon| \leq 1\%$ in our experiments) and the partial water filling of the cell is not felt [43]. Thus the mechanical properties in our compression tests should be dominated by the elastic bending of cellular walls, either faces or edges.

Secondly we discuss the question of why the root tissue in our experiments might behave like an open-cell foam. *A priori* this is not easy to understand for the root plant cells, especially for the ones of the cortex with solid walls spanning the faces of the polyhedral cell (like in **Fig. 13c**). However in [43], it was mentioned that the open or closed denomination for foams refers to the elastic rather than the geometrical character of the solid foam. It is possible that the root drying process in air modifies the relative importance of water pools and pathways (apoplastic, symplastic and transmembrane pathways) along the root section. The drying might result in a different distribution of subcellular water [45], for example in the junctions between cells, and modify the relative importance of cell wall faces with respect to cell edges for elastic properties. Another argument in favor of an open-cell foam model is that the cell wall is permeable and that the timescales involves in the transfer of water from cell to cell remains small compared to the cycling times at the strain rates we performed. It is also worthy to note that drying in air is not equivalent to drying in hypertonic solutions. Yet the effect of aging and long-term storage of plant tissues is frequently investigated in food science by immersing the tissue in hypertonic solutions of increasing concentrations. However, according to [36], "plasmolysis does not occur when a plant cell desiccates in air, because the water held by capillary forces in the cell walls causes the plasma membrane to remain pressed against the cell wall, even as the protoplast loses volume". The authors continue by arguing that "the cell (cytoplasm + wall) shrinks as a unit, resulting in the cell wall being mechanically deformed." This could suggest a possible mechanism of drying for the roots we mechanically tested in air and help understanding the experimental scaling law (behaviour analogous to dry open-cell foam) that we observed.

As a final comment, we should also recall that the fragment of chick-pea is cut at least 1 cm above the apex, meaning outside the meristematic and elongation zones of the root. Therefore the cells of the tested roots are in the differentiation zone with their length $h$ along the axis of compression longer than their width $a$. According to [46], this modifies the scaling laws relating the Young's modulus to the microscopic dimensions of the plant cell by introducing a pre-factor dependency with the corresponding aspect ratio $h/a$. Note that our experimental results are also in accordance with observations of [33]. In his work on shoots, the elastic modulus of a tissue increases, as the ratio of apoplast (cell walls) to symplast (protoplasm) increases. In our case, the drying of the whole material leads to a decrease of the root radius with time. If we assume that water is mainly lost in the protoplasm of cells in a first stage, the drying is associated to a densification of the tissue, resulting in an increasing proportion of the cell wall along the root section.

Mechanical properties at long drying times

A second regime is observed in our experiments for long times of drying with a departure from the power-law. This might be associated to water loss in the solid part, cell wall or middle lamella, thus resulting in a decrease of the total wall thickness (face or edge) between cells and a modification in the mechanical and also chemical properties of the cell wall. Water is known to be critical for the flow-like behaviour of cell wall matrix polymers [47]. Therefore the hypotheses made for establishing equation (11), i.e. the solid part keeps its volume as well as its mechanical properties, are no more valid. The primary cell walls of plants and vegetables are polymeric composite materials, consisting of a relatively amorphous matrix and a highly structured network of microfibrils embedded in the cell wall matrix. Upon intense drying, it is probable that the hygroscopic matrix polymers between the cellulose fibrils shrink [48], leading to an increase of the volume fraction of microfibrils of cellulose in the cell wall with possible induced anisotropy or rearrangement of microfibril orientation [49]. Yet the Young's modulus of the crystalline part of cellulose is known to be in the range of 130 GPa [50, 34], many orders of magnitude greater than the one of the polymeric gel [51]. Therefore the Young's modulus of the cell wall will increase upon intense drying, and thus the global Young's modulus of the tissue. The severe drying will also induce changes in the typical distance between cellulose microfibrils and certainly modify the complex chemistry of the cell wall [52] and therefore its extensibility [53], but these chemical effects upon drying are far out of the scope of this paper.

Finally this work might provide a possible alternative or complementary explanation for the observed decaying power relationship relating Young's modulus (and also ultimate strength) to root's diameters in soil science tests. In our experiments, the characteristic timescale for drying in air is smaller for thinner roots, meaning that roots of smaller diameter dry more rapidly. Consequently their Young's moduli increase faster with time. When testing roots of different diameters, this could have a drastic effect on the mechanical results, because the cutting and fixation of roots in a mechanical setup require some incompressible time delay during which the roots can dry if directly exposed to air. If the roots have initially an identical turgor pressure, the roots tested after a given time delay will give a higher Young's modulus when thinner, compared with thicker ones.

**Conclusions**

We performed mechanical characterization of roots drying in air. We observed a dramatic increase of the longitudinal root's Young's modulus with time when the root was exposed to air, but almost no variation for roots placed in osmotic solutions at the isotonic concentration to avoid water exchange and root's drying. The longitudinal Young's modulus of a compressed root in air was observed to scale as a decaying power-law with its section measured at different times of drying. We interpreted our results in the framework of the mechanics of cellular foams. This approach might rationalize some mechanical data obtained in soil science and explain the huge variability of Young's modulus and strengths of roots or shoots reported in literature. In some cases, this variability could be due to an evolution of the root moisture content, which is very sensitive to air drying, even for small time exposures. Thus the water status of the root has to be controlled during mechanical tests for obtaining reproducible data, which are crucial to implement models for soil stability or to understand the penetration of roots in soils.

**Acknowledgments**

We acknowledge Matteo Ciccotti and Eric Badel for very fruitful scientific discussions as well as Damien Vandembroucq for careful reading of the manuscript and Miguel Trejo for helping us in some of the experiments. We also want to thank all the members of the ROSOM project through the program of Agropolis Foundation (reference ID 1202-073) (Labex Agro: ANR-10-LABX-001-01). Ramon Peralta y Fabi is grateful for a Sabbatical Fellowship from DGAPA-UNAM.